\documentclass[submission, PhysCodeb]{SciPost}

\hypersetup{
    colorlinks,
    linkcolor={red!50!blue},
    citecolor={blue!50!red},
    urlcolor={blue!80!blue}
}

\usepackage[bitstream-charter]{mathdesign}
\usepackage{comment}

\DeclareSymbolFont{usualmathcal}{OMS}{cmsy}{m}{n}
\DeclareSymbolFontAlphabet{\mathcal}{usualmathcal}

\usepackage[color=orange!60,textsize=scriptsize]{todonotes}
\newcommand{\lincegenospace}[0]{Lincege.jl}
\newcommand{\lincege}[0]{Lincege.jl\ }

\usepackage{graphicx}
\usepackage{dcolumn}
\usepackage{bm}
\usepackage{listings}
\usepackage{minted}
\usepackage{amsmath}
\usepackage{caption}
\usepackage{subcaption}
\usepackage{amsthm}
\usepackage{svg}
\usepackage{algorithm}
\usepackage{algpseudocode}
\usepackage[english]{babel}
\usepackage[toc,page]{appendix}
\usepackage{booktabs}
\usepackage{authblk}

\title{\lincegenospace: LINked Cluster Expansions on a GEneral GEometry}

\author[1,2]{Pranav Seetharaman}
\author[1]{Francisco Correia}
\author[1]{Ehsan Khatami}
\affil[1]{Department of Physics and Astronomy, San Jos\'e State University, San Jos\'e, California 95192, USA}
\affil[2]{Department of Physics and Astronomy, University of Waterloo, Ontario, N2L 3G1, Canada}

\begin{document}

\maketitle

\begin{abstract}
The numerical linked-cluster expansion (NLCE) is a powerful tool for studying properties of quantum lattice models directly in the thermodynamic limit. Here, we introduce \href{https://github.com/rightleftspin/LiNCEGE.jl}{\lincegenospace}, a versatile, efficient, and easy-to-use package for generating the clusters needed to perform the NLCE on a general finite or infinite lattice. \lincege takes as input only the Bravais lattice vectors, the basis information, and the expansion unit and uses a depth-first search to efficiently generate clusters, their lattice constants, and the sub-cluster information. It outputs final coefficients of the series, expressed in terms of the extensive properties of the clusters, in different orders. The package also contains tools for visualizing the lattice and the generated clusters as well as a classical Ising model solver that can be used as a template for interfacing \lincege with solvers of other lattice models. This package
can be installed using the Julia package manager and utilized with example scripts from our GitHub repository at \url{https://github.com/rightleftspin/Lincege.jl}. The online documentation, including examples, can be found at \url{https://rightleftspin.github.io/Lincege.jl/stable/}.
\end{abstract}

\tableofcontents

\section{Background and motivation}
The numerical linked-cluster expansion (NLCE), first introduced twenty years ago through a series of works by Rigol, Bryant, and Singh~\cite{rigol_numerical_2006,rigol_numerical_2007,rigol_numerical_2007b}, is a powerful technique for obtaining finite-temperature properties of quantum lattice models in the thermodynamic limit. It utilizes an expansion in terms of exact contributions from finite-clusters on the lattice. The formulation, which is discussed in some detail in Sec.~\ref{sec:formulation}, resembles that of high-temperature series expansions, although it lacks an explicit small parameter, like the inverse temperature. Its original applications were to quantum magnets on different two-dimensional geometries and their thermodynamic properties~\cite{rigol_numerical_2006,rigol_numerical_2007,PhysRevLett.98.207204,PhysRevB.76.184403}. However, it was found to be an effective method for the study of itinerant electron models, such as the $t-J$ and Hubbard models~\cite{rigol_numerical_2007b,PhysRevA.84.053611,PhysRevA.86.023633}. Its comparable performance to other unbiased state-of-the-art finite-temperature methods for strongly-correlated electron models, such as the determinant quantum Monte Carlo (DQMC)~\cite{r_blankenbecler_81}, coupled with its access to the exact partition function and all the other thermodynamic properties on arbitrarily fine temperature and chemical potential grids in relatively fast execution times, made it a reliable method for theory verification of experiments with ultracold atoms in optical lattices~\cite{r_hart_15 ,PhysRevLett.114.070403,l_cheuk_16,p_brown_17,m_nichols_19,PhysRevLett.132.083401,lebrat_observation_2024,mongkolkiattichai2025}. In fact, the NLCE was found to be complementary to the DQMC in that its convergence improves as, for example, the on-site interaction in the Hubbard model increases and the system approaches the atomic limit, while the DQMC generally performs best in the weakly-interacting region of the model (see e.g., Refs.~\cite{PhysRevLett.109.205301,PhysRevB.91.241107}). Later, the NLCE was also used to study entanglement~\cite{PhysRevB.84.165134,PhysRevB.86.075106,PhysRevLett.110.135702,Kallin_2014,PhysRevB.93.085120}, three-dimensional systems~\cite{khatami_three-dimensional_2016,schafer_pyrochlore_2020}, nonequilibrium dynamics after a quench~\cite{PhysRevE.95.033302,PhysRevLett.120.070603,PhysRevX.9.021027,white2021}, inhomogeneous and disordered systems~\cite{PhysRevB.91.174413,PhysRevB.99.205113,PhysRevA.102.013318,PhysRevE.109.054127}, SU(N) Hubbard models~\cite{PhysRevA.104.043316,PhysRevLett.132.083401,PhysRevA.108.053312,mongkolkiattichai2025}, and dynamical properties~\cite{m_nichols_19,PhysRevB.99.094419,pham2026}.

So far, to apply the technique to a new lattice geometry or Hamiltonian, one has had to develop new cluster generation and analysis codes, in most cases from scratch. A unified framework for dealing with generic lattice models has been missing. The issue of versatility with the NLCE applications has motivated us to develop a package in Julia we call LINked Cluster Expansions on a GEneral GEometry, or {\it Lincege} for short. 

In \lincege, we take a systematic approach and do not assume a particular geometry or even Hamiltonian. Instead, the user can specify the Bravais lattice vectors and the locations of the sites in a unit cell in the case of regular lattices, or the bond and other information of a large enough lattice in case of an irregular structure, in order to obtain clusters and all the other information needed to carry out the NLCE. Moreover, the calculations are not necessarily limited to the thermodynamic limit and can be carried out for finite clusters. Therefore, we believe that this package presents a major advancement with regard to this important tool for studying finite-temperature properties of quantum lattice models, especially in the thermodynamic limit.

We have designed \lincege to offer minimal interaction with the code for a generic user interested only in obtaining clusters of a well-defined and typical quantum lattice problem while at the same time have built in capabilities to make add-ons to the code or generalizations rather straightforward and user-friendly. We have developed and included in the package several examples, which we discuss in detail below. All the codes are open-source and documented.

The new approach, including the development in Julia, has allowed us to take advantage of already available and extremely efficient packages when it comes to the bottlenecks of the algorithm. For example, the isomorphism check, which usually takes the vast majority of the time when it comes to the generation of topologically distinct clusters and their sub-cluster information has been outsourced to \textit{nauty}~\cite{mckay_practical_2014}, an algorithm that efficiently finds the canonical representative for the given cluster. The main improvement comes from the usage of sequential graph hashing, which represents clusters as easily comparable hashes based on the underlying translations, symmetries, or isomorphisms. This allows us to rethink the order of operations traditionally done in NLCE algorithms so that one can skip intermediary classification steps such as symmetric classification if it is not needed explicitly.

\section{Formulation of the Method}
\label{sec:formulation}

Linked-cluster expansions (LCE)~\cite{Domb,linked} are based on the idea that any extensive property of a lattice model, $P$, can be expressed as a series in terms of contributions from all clusters, $c$, that can be embedded in the lattice: 
\begin{equation}
    P(\mathcal{L}) = \sum_c W_p(c),
    \label{eq:LCE}
\end{equation}
where $\mathcal{L}$ represents the lattice, and $W_p(c)$ is the {\it weight} or the {\it reduced property} for cluster $c$. This weight is calculated through the inclusion-exclusion principle by computing the property for the cluster, $P(c)$, and subtracting from it the weights of all the smaller clusters, $s$, that can be embedded in $c$ ({\it sub-clusters} of $c$), i.e., 
\begin{equation}
    W(c)=P(c)-\sum_{s\subset c}W(s).
    \label{eq:W}
\end{equation}
This is a recursive process which often starts with $W(c)=P(c)$ for $c$ being a single site. In the numerical linked-cluster expansion (NLCE)~\cite{rigol_numerical_2006,tang_short_2013}, $P(c)$ are computed using exact numerical methods such as the exact diagonalization, quantum Monte Carlo, density matrix renormalization group, etc.

It is easy to see that Eqs.~(\ref{eq:LCE}) and (\ref{eq:W}) are in fact the same equation; by separating the weight of the property for the lattice itself, $W(\mathcal{L})$, from the sum in Eq.~(\ref{eq:LCE}), one can arrive at Eq.~(\ref{eq:W}). However, the series in Eq.~(\ref{eq:LCE}) is often written for the lattice in the thermodynamic limit, or for a very large lattice beyond the reach of exact methods, in which case it has to be truncated at some point. Moreover, it would be practical to work with the property per site in that case, i.e.,
\begin{equation}
    P(\mathcal{L})/N = \sum_c L(c)W_p(c),
    \label{eq:TL}
\end{equation}
where $N$ is the number of sites (which can be infinity) and $L(c)$ is the embedding factor, or the {\it lattice constant}, defined as the number of ways {\it per site} the now distinct cluster $c$ can be embedded in $\mathcal{L}$. For example, if the lattice has translational symmetry, it suffices to consider only those clusters that are not related by a translation operation. In practice, clusters that are related through other symmetry operations or otherwise have the same weight may be bundled together in this series with their lattice constants added up to reduce the cost of computations. In Sec.~\ref{sec:anatomy}, we discuss the implementation of such checks in \lincegenospace. Note that the series in Eq.~(\ref{eq:TL}) can ultimately be written entirely in terms of $P(c)$, albeit with coefficients that are linear combinations of $L(s)$ for $s \subseteq c$.

\section{Anatomy of the Algorithm}
\label{sec:anatomy}
\begin{figure}[h]
    \centering
    \includegraphics[width=0.9\linewidth]{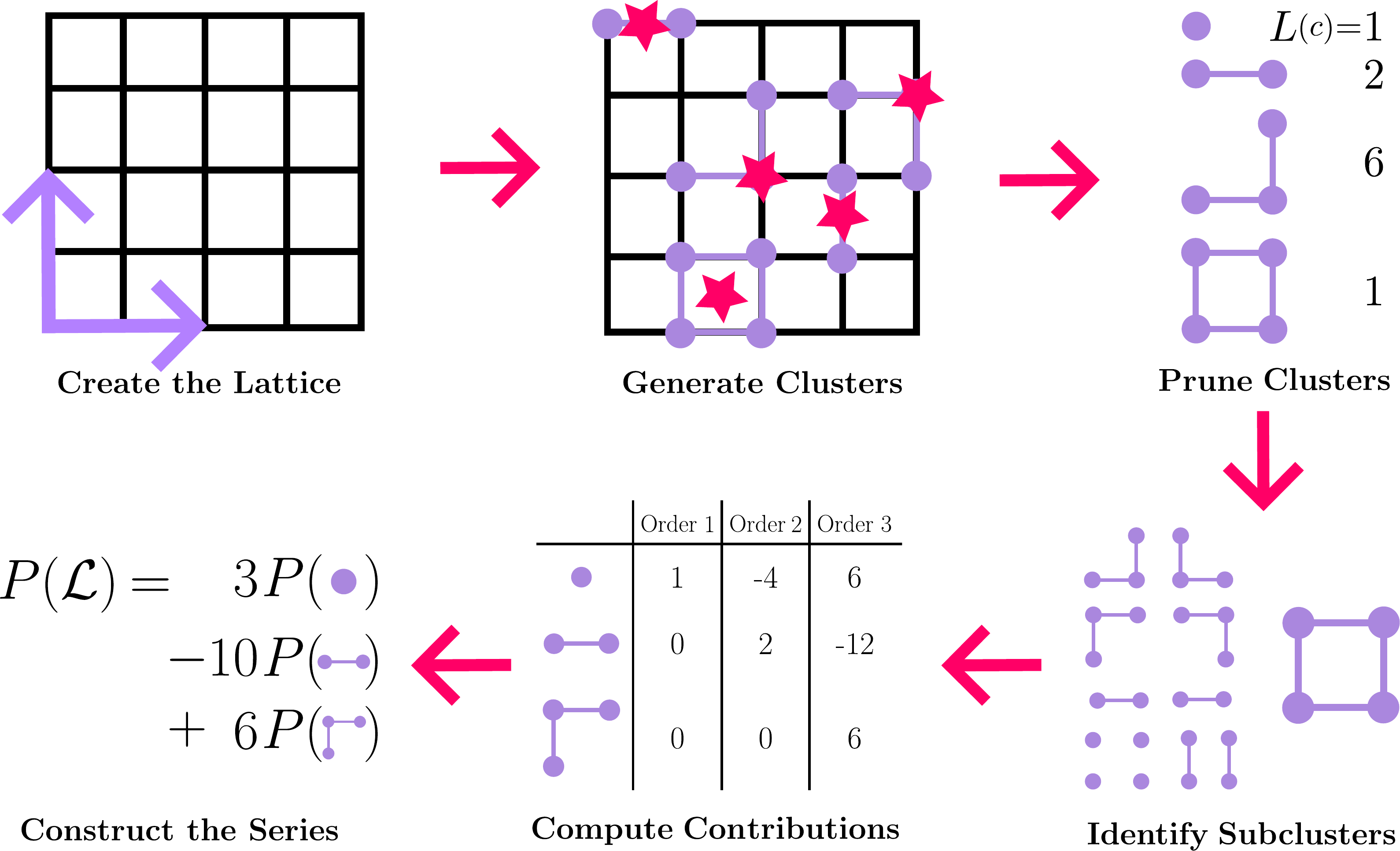}
    \caption{Details the LCE process as performed by \lincegenospace. The first step takes in the lattice information and generates the lattice and appropriate bond information. Step two generates clusters on the lattice: For infinite lattices, \lincege starts from the center point. For finite lattices, every site is explored. Step three prunes the generated clusters by some equivalence relation. The example shows equivalence by graph isomorphism. \lincege then identifies the subclusters of every cluster and collects them based on the pruning equivalence relation. Then, \lincege uses this subcluster information to compute the contributions from each cluster at each order (the table in the middle of the bottom row). Note that these are per order corrections (corresponding to {\em partial sums}) and the total coefficient along each row needs to be calculated for the final step where the LCE sum is computed. In the lower left corner, we show the series in order 3.}
\label{fig:nlce_flowchart} 
\end{figure}

\lincege is a collection of algorithms that work together to (i) generate clusters, (ii) collect them by equivalence relations, and (iii) compute their appropriate coefficients for the LCE. The workflow is shown in Fig.~\ref{fig:nlce_flowchart} for the example of the {\em site expansion} on the square lattice, in which the expansion unit is a single site and the expansion {\em order} is the number of sites in the largest clusters. In this section, we elaborate on the algorithms used in \lincege to implement these three steps. For readability, we first discuss equivalence relations between clusters instead of cluster generation.

\subsection{Equivalence Relations and Cluster Hashing}\label{subsec:equiv}
\par A key part of NLCE is collecting clusters together based on shared properties. Relating clusters that have the same Hamiltonian (or other observables) despite having different forms significantly reduces computation when performing exact diagonalization. To partition sets of clusters, \lincege employs a suite of equivalence relations realized by cluster hashing functions.

\subsubsection{Translational Equivalence}\label{subsubsec:translational_equiv}
\par A useful way to compare clusters is to see if one cluster can be translated onto another by a translation that preserves the underlying lattice. Identifying translationally equivalent clusters are a necessary first step for infinite lattices since they have a known lattice constant of $1/n_T$ where $n_T$ is the number of sites that are not related by translations in the translation set of the lattice. The set of all translations that preserve the lattice is denoted as $\mathcal{T}$, where an individual translation $T \in \mathcal{T}$ acts on a cluster $c$, translating it to another cluster $c' = T\cdot c$. We can then define an equivalence relation, $\sim_\mathcal{T}$, which equates two clusters $c_1$ and $c_2$ if there is a lattice translation $T \in \mathcal{T}$ such that $T\cdot c_1 = c_2$. Note that some clusters may be possible to translate onto one another through some translation not in $\mathcal{T}$, but these are not considered equivalent under $\sim_\mathcal{T}$.

\par
\begin{figure}[!h]
    \centering
    \includegraphics[width=0.7\linewidth]{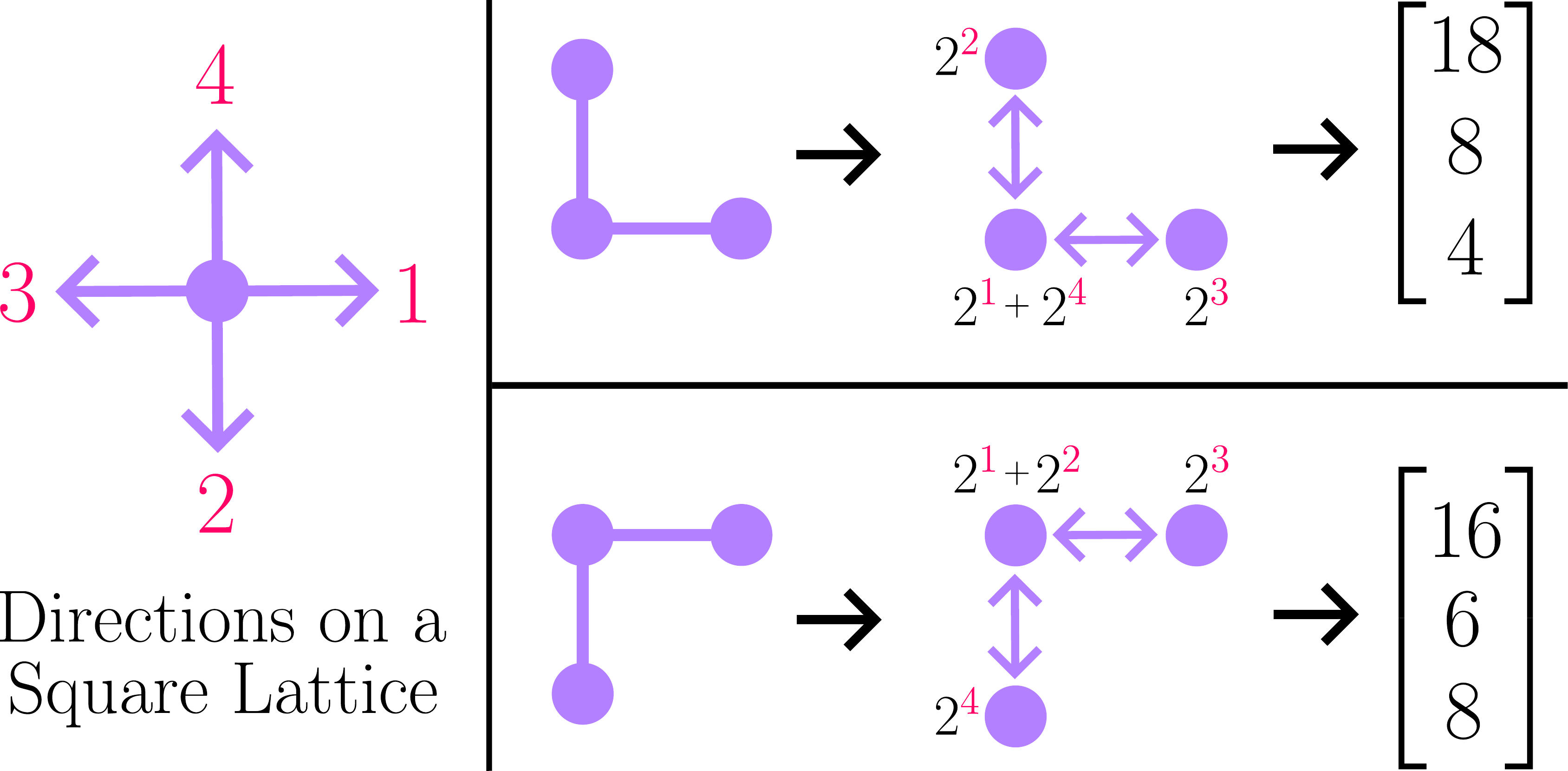}
    \caption{Example of two clusters on a square lattice that are not translationally equivalent. The minimum vertex indexing scheme is followed and a translationally invariant representation is generated for each cluster. In this case, the consistent path is read from left to right, bottom to top.}
    \label{fig:MVI}
\end{figure}

 In practice, determination of translational equivalence between two clusters is a difficult problem. Relating two clusters by some translation in a set requires applying each translation to one cluster and checking for coordinate equivalence. Instead, \lincege encodes translational equivalence through a hashing scheme that is a modified form of minimum vertex-type sequence indexing (MVI) as presented by Liao {\it et al.} in Ref.~\cite{liao_minimum_2017}. MVI is a process by which a cluster on an underlying lattice can be represented by a vector of integers corresponding to the directions of each bond per vertex. A visual representation of MVI is provided in Fig. \ref{fig:MVI} where we show two clusters that cannot be translated onto each other and their corresponding MVI vectors. Each vertex is labeled by the directions of the bonds around it and the MVI vector is ordered by following a consistent path through the lattice. In our case, we sort the vertices in the lattice from left to right and bottom to top consistently among all the different clusters and the MVI vector is formed in this order. The two MVI vectors can now be compared to ensure translational equivalence.

\subsubsection{Isomorphic Equivalence}\label{subsubsec:isomorphic_equiv}
\par For many physical models, we can choose a coarser partition than translational equivalence to reduce redundant computations. The Hamiltonians of translationally invariant quantum lattice models, for example, depend only on the bond information of a cluster and not necessarily on the location or relative displacements of its vertices. With this in mind, we can partition the set of translationally distinct representative clusters into collections of isomorphically equivalent clusters. We can once again choose to work with representatives of each member of the partition for the solution of the Hamiltonian, but the lattice constant of the representative is now the sum of all the lattice constants of each cluster in the partition. This is why it is useful to start with clusters for which the lattice constants are already known from the generation step. Finding lattice constants after isomorphic partitioning would not be an efficient process.

\par \textit{nauty}~\cite{mckay_practical_2014}, \textit{bliss}~\cite{JunttilaKaski:ALENEX2007}, and SAUCY~\cite{Darga2004ExploitingSI} are examples of the existing algorithms that determine if two graphs are isomorphically equivalent. However, this problem is notoriously difficult. Furthermore, we aim to determine uniqueness in a list of graphs, thus finding equivalence between two graphs is not enough, hashing is necessary. Finding an isomorphic hash for a graph requires finding a unique label for isomorphically equivalent graphs. This canonical labeling is very efficiently done by \textit{nauty}, a graph isomorphism package. \textit{nauty} combs through every unique labeling of a given vertex-colored graph to find the canonical labeling~\cite{mckay_practical_2014}. \lincege uses \textit{nauty} with an additional wrapper we have utilized that works for edge colored graphs. With the canonical labeling in hand, \lincege can partition the translationally distinct clusters further for lowered computational load.

\subsubsection{Symmetric Equivalence}\label{subsubsec:symmetric_equiv}
\par Another coarser partition one can choose is symmetric equivalence. Two clusters are considered equivalent if some symmetry of the lattice relates them. For certain observables, such as correlation functions, we need to differentiate between clusters that are isomorphically equivalent when it comes to the Hamiltonian, but have different relative bond orientations and therefore different Euclidean distances between vertices on the lattice, making symmetric equivalence relevant. Practically, determination of symmetric equivalence involves applying all point-group symmetries of the lattice to a cluster then centering the cluster and checking against all the other clusters. This approach is not computationally scalable, so we instead use a unique hashing scheme based on a canonical representative of the minimum vertex indexing of every symmetric-transformed form of the cluster. 
\par When creating a lattice, \lincege takes the set of symmetries and converts them into permutations. These permutations are then applied to every cluster (only translationally invariant clusters in the case of the infinite lattice) and a canonical representative is chosen from the symmetric orbit of that cluster. This representative is now the hash for confirming symmetric equivalence among clusters. While this procedure is optimized, the isomorphic hashing scheme is much faster. Thus, in \lincege, we only utilize symmetric hashing when the user requests it. It is much faster to utilize symmetric hashing only when needed and not as a step towards collecting isomorphic clusters.

\subsection{Generation of Clusters}\label{subsec:ClusterGeneration}
\par For a given lattice, the first step \lincege performs is the generation of every cluster that can be embedded on the lattice. The benefit of this approach is that the lattice constant for each cluster generated this way will be $1/N$, where $N$ is the number of sites in the lattice, since every cluster is thought of as unique. For a finite lattice, this approach is computationally feasible, since all clusters can be enumerated from the lattice in a finite amount of time. However, an infinite lattice renders this approach computationally impossible since it would require generating an infinite number of clusters. 

\par Thus, our new goal for a LCE on an infinite lattice is to reduce the infinite set, $S$, of linked clusters embedded on the lattice to a finite set that is possible to compute. With this in mind, we can choose to partition $S$ by equivalence under the translations $\mathcal{T}$ that preserve the infinite lattice. We then define a partition $P = S\ /\sim_\mathcal{T}\ = \{m_s: s \in S\}$, where $P$ is a collection of member sets denoted as $m_s = [s] = \{g \sim_\mathcal{T} s: g \in S \}$. Each member $m_s$ is a set containing all clusters that are equivalent under lattice translations of the arbitrary representative cluster $s$. Conveniently, the size of any member $m_s \in P$ is equivalent to the number of unit cells in the lattice~\footnote{Note that this statement is dependent on the chosen expansion unit. An expansion unit that breaks the translational symmetry of the underlying lattice such as a corner sharing square expansion on the square lattice requires more care.}, which is represented as a finite lattice on the computer. Thus, the lattice constant can be defined for a representative $s$ as $L(s) = \frac{1}{n}$ where $n$ is the number of sites in the unit cell. \lincege can then generate the representative clusters of each member of $P$ since the set of representatives is finite at a given order and the lattice constant for each representative cluster is known.

\par
\begin{figure}[!h]
    \centering
    \includegraphics[width=0.9\linewidth]{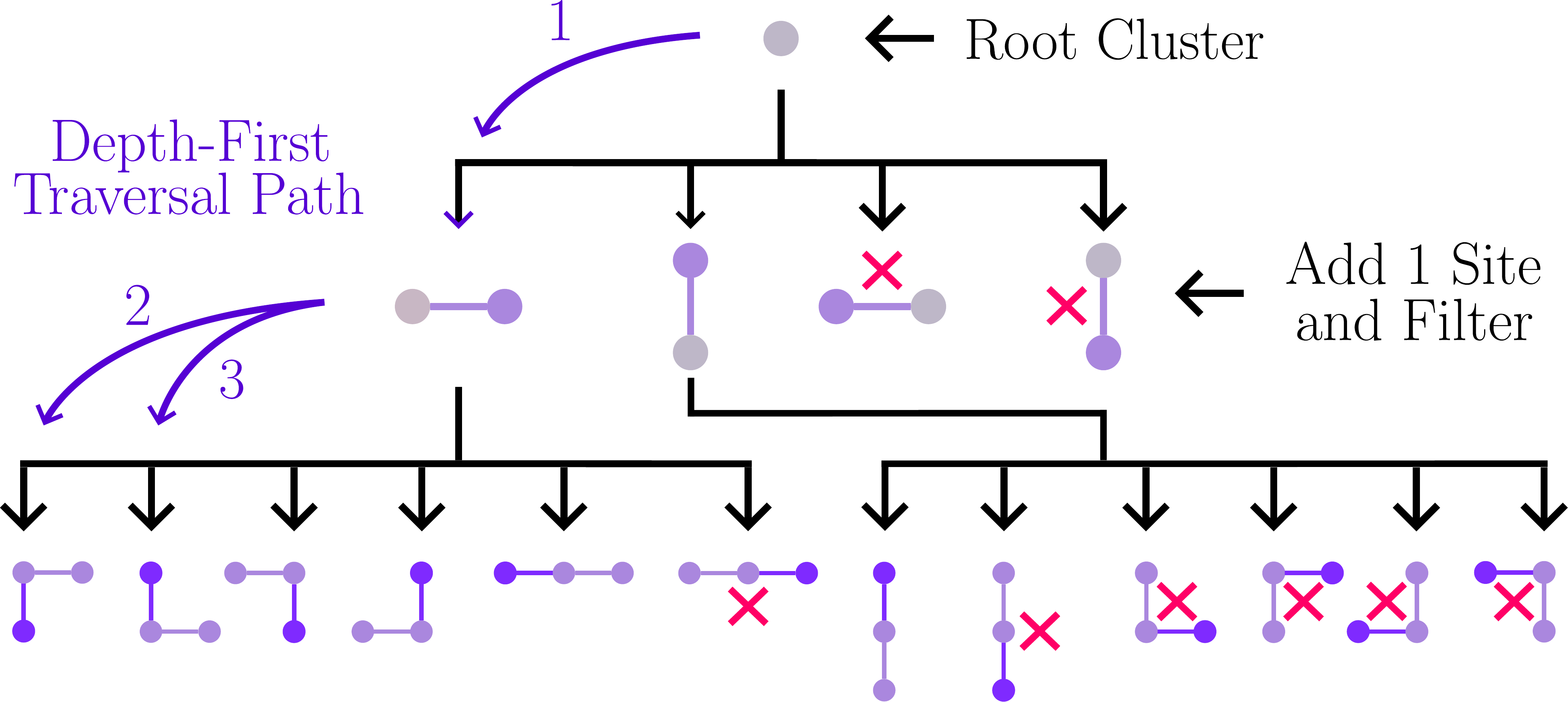}
    \caption{Visualization of a depth-first enumeration of the tree of all possible subgraphs up untill order 3. The path taken by the algorithm is numbered for the first three steps. These numbers are unrelated to translational directions shown in Fig.~\ref{fig:MVI}. The crossed out clusters are equivalent by a lattice translation so as to generate a set of translationally distinct clusters.}
    \label{fig:dfe}
\end{figure}

 \lincege represents the set of all clusters that can be embedded on a lattice through a tree graph. The root node of the tree is a single tiling unit of the lattice, and the downward neighbors of each element are every cluster that can be reached by adding a single tiling unit to the cluster. We can see a visualization of this tree up to order 3 for a square lattice when the tiling unit is a single site on the left hand side of Fig. \ref{fig:dfe}. To explore this tree for both finite and infinite lattices, \lincege uses a parallelized, depth-first enumeration of the tree for a given lattice. The tree spans all possible connected clusters, but \lincege prunes this tree down to translationally distinct clusters for an infinite lattice, as shown by the redundant, crossed out clusters in Fig. \ref{fig:dfe}. This is a depth-first search with the set equivalence relation interchangeable based on the type of lattice.

\par \lincege collects all clusters in the tree that are unique with respect to a given hashing function. For a finite lattice, this hashing function distinguishes every cluster with different lattice sites as unique. For an infinite lattice, the hashing function represents each cluster in a form that is invariant under the lattice translations~\cite{liao_minimum_2017}. \lincege has the ability to accept custom hashing functions in the depth-first traversal.

\subsection{Sub-Cluster Coefficients and Final Cluster Coefficients}\label{subsec:ClusterCoeff}
The final step of any linked-cluster expansion is to find subclusters of each enumerated cluster and compute their appropriate coefficients at each order. This process is effectively the same as for the lattice, but done for each cluster. Similar to the depth-first traversal of the tree of all clusters on the lattice, \lincege performs a depth-first traversal of all subclusters of a given cluster. This depth-first traversal returns subclusters and their cluster constants, which count the number of ways a given subcluster can be embedded into the cluster.
\par Importantly, \lincege reports final cluster coefficients of properties as per order corrections after the NLCE recursive summation has been performed to obtain the weights $W_p(c)$ in terms of $p(s \subseteq c)$. This removes the need to perform the summation numerically for each property separately. Table \ref{tab:square_final_example} shows the counts for the example of the square lattice site expansion up to order 3.

\begin{table}[!h]
 \centering
\begin{tabular}{ccccr}
  \hline
  Order & Single Site Coeff. & Dimer Coeff. & Trimer Coeff. & $P(\mathcal{L})/N$ \\
  \hline
  1 & 1 & 0 & 0 & $P(s)$\\
  2 & -4 & 2 & 0 & $-3P(s) + 2P(d)$\\
  3 & 6 & -12 & 6 & $3P(s) - 10P(d) + 6P(t)$\\
  \hline
\end{tabular}
\caption{Output coefficients of properties of individual clusters from \lincege for partial sums in each order (middle three columns) and how they need to be combined to obtain the final sum of the NLCE at each order, including the lower order contributions (rightmost column). The results are shown up to order 3 for the site expansion on the square lattice. The three isomorphically distinct clusters are the single site, the dimer and the trimer, the properties of which are denoted $P(s)$, $P(d)$, and $P(t)$ respectively.}
\label{tab:square_final_example}
\end{table}

We have built into \lincege an exact enumeration solver for the classical Ising model, which demonstrates how the output of \lincege can be used in conjunction with a cluster solver that provides $p(c)$ in practice. It can serve as a starting point for studying other Hamiltonians and  integrating other solvers into the workflow.

\subsection{Cluster Expansions}\label{sec:cluster_expansions}
The convergence of an NLCE algorithm can be altered via the choice of the fundamental building block. So far, we have mainly discussed building up the lattice via single sites, known as a site expansion. However, one could imagine building up a lattice via larger building blocks, such as a square lattice via corner sharing squares. This is known as a weak cluster expansion as the tiling units share sites. Similarly, we can perform a unit cell expansion, e.g., using up-triangles on the kagome lattice. Since the cells here are disjoint, this is considered a strong cluster expansion. \lincege supports these types of expansions by treating the center of each expansion unit as a site in a site expansion. It then keeps track of corresponding bonds and sites and switches between the expansion lattice made of the centers of each expansion unit and the physical lattice made of the physical sites based on the use case. Hashing and cluster identification are always done on the physical lattice to maintain correctness. Generated tables for cluster expansions exist in Appendix \ref{appendix:counts}, and detailed code examples can be found on the documentation website.

\subsection{Performance}
\par For quantum lattice models, the true time limitation for NLCE is often the exact evaluation of various physical observables on individual clusters in the series, for which exact diagonalization is typically used. The generation of NLCE clusters and their information is something that needs to be done only once for any model and should be a relatively quick endeavor. Nevertheless, we acknowledge and rectify a few bottlenecks in \lincege inherent to the cluster generation pipeline in Sec.~\ref{sec:bottlenecks}.

\par We note that as the order of the expansion increases, the coefficients of $p(c)$, especially for clusters in lower orders, can grow very large in magnitude and change sign from one order to the next. The final NLCE series effectively reduces to a subtraction of two numbers, whose relative difference in magnitude decreases rapidly. As a result, any small uncertainty in properties, such as a systematic or statistical error, especially for clusters in low orders, can propagate quickly and render the NLCE useless. Nevertheless, methods other than exact diagonalization can be employed to solve the clusters depending on the problem~\cite{Kallin_2014}.

\subsection{Major Algorithm Bottlenecks}
\label{sec:bottlenecks}
\par We have tackled three bottlenecks for an efficient implementation of the LCE cluster generation pipeline. These are the depth-first search within the cluster generation routine, the cluster hashing routine within the cluster uniqueness determination, and the generation of subgraphs for each cluster within the weights generation routine.
\par The easiest to tackle has been the the cluster hashing routine, where the problem is straightforward to parallelize. One can simply compute the expensive hash for each cluster in parallel and collect equivalent clusters sequentially. The next easiest task to tackle has been the generation of subgraphs for each cluster, which is directly parallelized for each cluster. Additionally, a thread specific cache is maintained to easily hash each subgraph from just the vertices instead of the expensive call to the corresponding hashing function.
\par The depth-first search is the most difficult to parallelize, but this is also the most important step to consider. The depth first search is the most time intensive part of any LCE routine. The generation of clusters is a computationally difficult task in both time and space complexity, and it is a notoriously difficult problem to parallelize effectively. \lincege achieves this parallelization through a shared thread-safe stack and multiple threads that access and modify the stack. Each thread is responsible for collecting new clusters from the shared stack, and adding new clusters up untill the appropriate order. This is not the most efficient algorithm, but presents a noticeable speed improvement in real-world use cases. Future improvements to this algorithm are planned, including the use of batch assignment for each thread and persistent threads instead of the currently used ephemeral threads.

\section{Performance}\label{sec:performance}
\begin{figure}[h]
    \centering
    \includegraphics[width=0.9\linewidth]{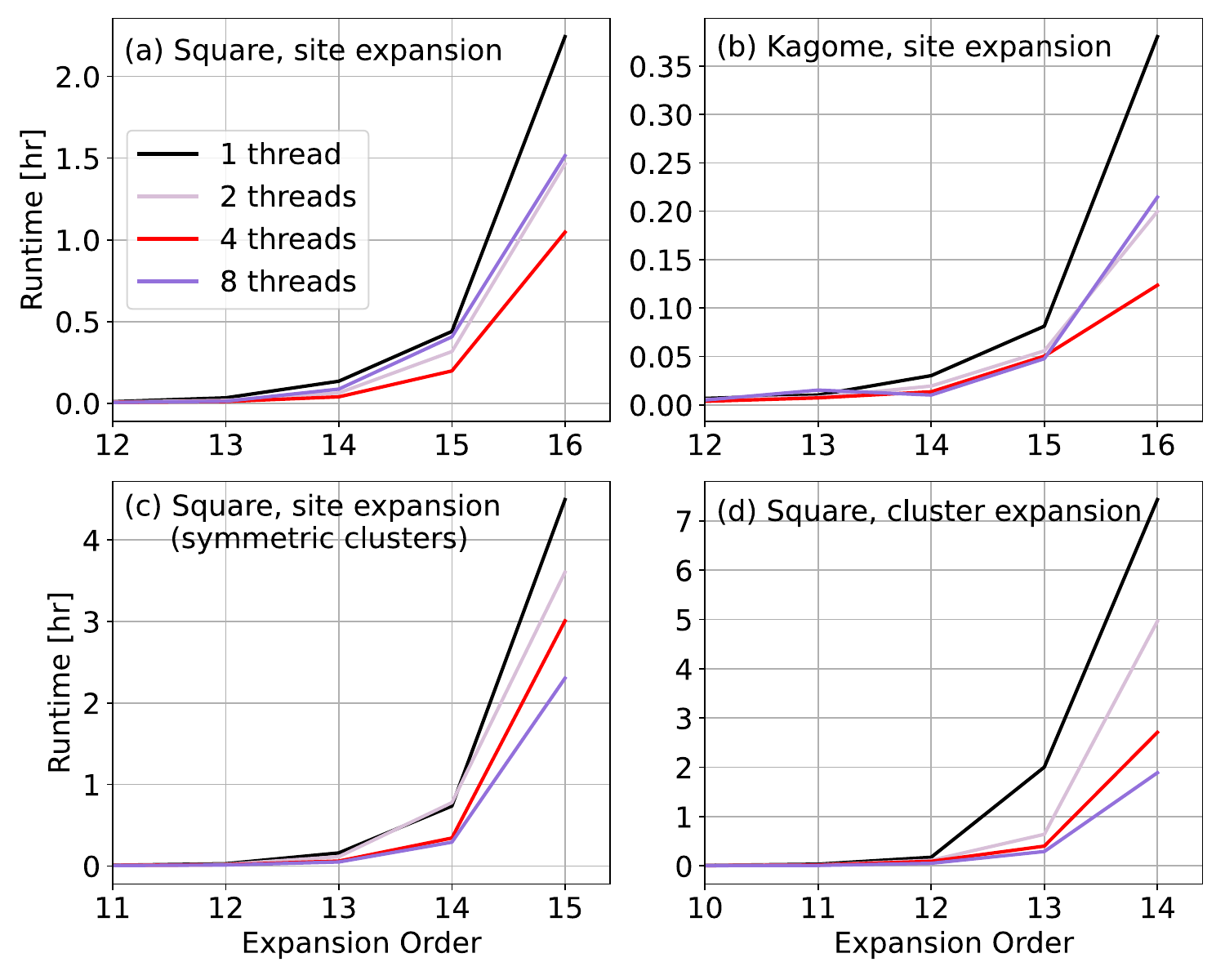}
    \caption{\lincege run times as a function of the expansion order for different expansion schemes. In (a), (b), and (d), only isomorphically distinct clusters are generated. In (c), all the symmetrically distinct clusters are generated.}
    \label{fig:bench}
\end{figure}

In Fig.~\ref{fig:bench}, we show run times of single-thread and multi-threaded \lincege for four different expansion schemes. We find that for site expansions, the largest gain, which is about a factor of two for the square lattice [Fig.~\ref{fig:bench}(a)] and a factor of three for the Kagome lattice [Fig.~\ref{fig:bench}(b)] in order 16, is with four threads. Further increasing the thread count to eight leads to a slowdown potentially due to the communication overhead. However, the site expansion on the square lattice in which all symmetrically distinct clusters are generated [Fig.~\ref{fig:bench}(c)] or the one in which a $2\times 2$ square is used as the expansion unit [Fig.~\ref{fig:bench}(d)] continue to gain a modest speedup by doubling the thread count to eight up to orders 15 and 14, respectively. The performance results below are measured on Intel CPUs with 2.40 GHz clock speed and 64 GB of RAM up to order 12 and then increasing to 256 GB of RAM up to order 16. 

\section{Summary and Outlook}
We have implemented an efficient, versatile, and user-friendly LINked Cluster Expansions on a GEneral GEometry (or Lincege for short) in Julia that can be utilized to perform NLCE calculations for quantum lattice models. The \lincege package represents the set of all clusters that can be embedded on a finite or infinite lattice through a tree graph, whose root node is a single expansion unit. It uses a parallelized depth-first enumeration of the tree to generate the clusters in the series, and it repeats the same process for every cluster in order to obtain its sub-clusters and their coefficients. \lincege then encodes translational equivalence through a hashing scheme that is a modified version of the minimum vertex-type sequence indexing algorithm. To make realistic NLCE calculations efficient, partitionings based on isomorphism or symmetric equivalence is done through canonical labeling and hashing of graphs, including on geometries for which there are more than one type of vertex or edge. \lincege can also be utilized to generate clusters for series with an expansion unit that contains more than one lattice site. In Appendix~\ref{appendix:examples}, we present several example inputs to \lincege for different geometries and expansion units, along with the generated clusters in the first few orders. For the latter, we utilize a visualization function we have included in \lincegenospace. More examples, driver codes, and the documentation can be found in the project page at \url{https://rightleftspin.github.io/Lincege.jl/stable/}.

Future versions of \lincege are planned, with performance and functionality being key motivators. The performance of the parallelized algorithms contained in \lincege could be improved if they were written in a programming language with a lower cost to spawning threads. These improvements would both improve speed and memory requirements, especially at higher orders. Additional functionality is also planned, including reporting of graph orbits to speed up exact diagonalization calculations with intelligent symmetry identification~\cite{schafer_pyrochlore_2020}. Additionally, improved documentation from real-world use is a central goal of the future updates to \lincegenospace.

\section{Acknowledgments}
The authors acknowledge helpful discussions with Robin Newby, Lorenzo Mambretti, Parthiv Seetharaman, and Jason Pruitt. This work was supported by the U.S. Department of Energy, Office of Science, Office of Basic Energy Sciences, under Award Number DE-SC0022311. F.C. acknowledges support from the National Science foundation’s (NSF) Research Traineeship (NRT) Fellowship funded by the Grant No. DGE-2125906. E.K. acknowledge support in part by grant NSF PHY-2309135 to the Kavli Institute for Theoretical Physics (KITP), where part of their work was performed. The source code for \lincege represents an original work by the authors of this paper (and the developers of \textit{nauty}, another package we utilize). No generative AI was used in the development of the core NLCE code or the writing of this manuscript.

\begin{appendices}
\section{Examples}
\label{appendix:examples}

The general workflow in the utilization of \lincege is the same for all geometries and consists of the following steps: (i) Define the primitive vectors and the basis for the expansion unit, (ii) indicate the basis site {\it types}, (iii) define the bonds of the lattice, and in case of cluster expansions, bonds of the lattice used for the expansion, (iv) indicate the bond types, and (v) indicate the expansion order.  In order to assess the accuracy of the above inputs, we have implemented imaging functions that can be used to visualize the lattice and generated clusters prior to the computation of their coefficients and subcluster information. Specific functions and notations may change in later versions, and so, we highly recommend consulting the most up to date documentation. In the following, we showcase the output of \lincegenospace's imaging tool after discussing the inputs for three sample geometries/expansion schemes in the order of complication: (1) The site expansion on the simple square lattice, (2) the $2\times 2$ cluster expansion on the square lattice, (3) the site expansion on the Kagome lattice, and (4) the site expansion an anisotropic Shastry-Sutherland lattice with different types of sites.

\subsection{Square Lattice}

\subsubsection{Site expansion}

\begin{figure}[h]
    \includegraphics[width=\linewidth]{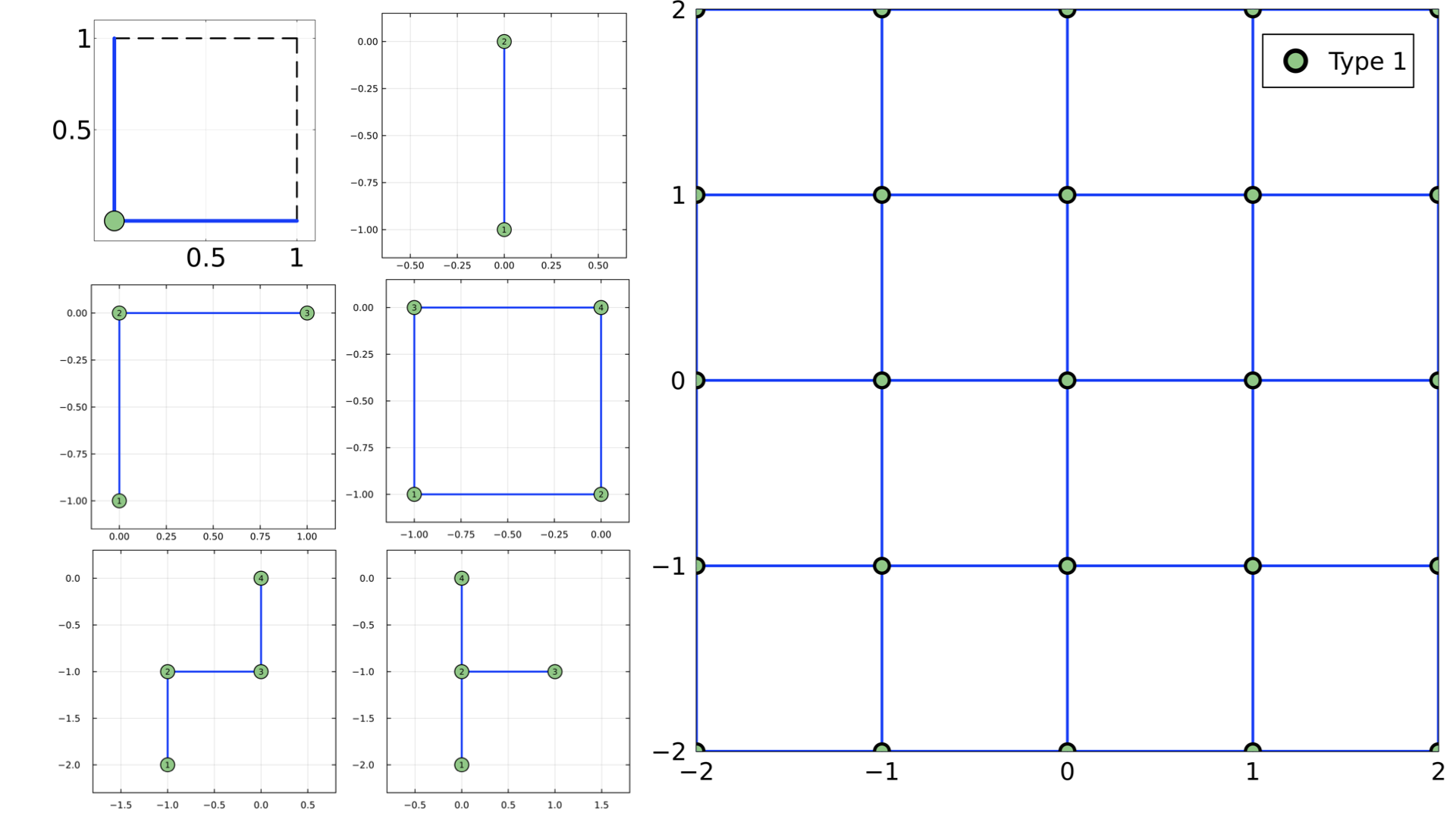}
    \caption{Right: The simple square lattice. Top left: The unit cell and primitive vectors. Left: All topologically unique clusters with up to 4 sites excluding the single site. 
    }
    \label{fig:delta}
\end{figure}

The input to \lincege for the site expansion on the simple square lattice with one bond type is presented in Table~\ref{tab:square_lattice}. Figure~\ref{fig:delta} visualizes the lattice (right), the unit cell and primitive vectors (top left), and isomorphically distinct clusters up to order four of the site expansion on this geometry.

\begin{table}[h]
\centering
\begin{tabular}{@{}lll@{}}
\toprule
\textbf{Parameter} & \textbf{Symbol} & \textbf{Value} \\
\midrule
Primitive Vectors 
  & $p_1$ & $[1,\,0]$ \\
  & $p_2$ & $[0,\,1]$ \\
\addlinespace
Basis Vector
  & $b_1$ & $[0,\,0]$ \\
\addlinespace
Basis Type 
  & $f$ & $[1]$ \\
\addlinespace
Lattice/Expansion Bonds
  & $L_1$ & $[1,\,1,\,[1,0],\,1]$ \\
  & $L_2$ & $[1,\,1,\,[0,1],\,1]$ \\
\bottomrule
\end{tabular}
\caption{The input for the site expansion on the square lattice with a single-atom basis and a uniform bond type. The basis along with the primitive vectors make the underlying Bravais lattice. Basis type(s) $f$ is a list of numbers indicating the type of sites in the basis. Each {\it lattice bond} (same as {\it expansion bond} in the site expansion -- a bond belonging to the expansion unit) contains, in order from left to right, the site index of the first site in the unit cell, the site index of the second site in the unit cell, coefficients of the translation vectors $p_1$ and $p_2$ required to get from the first site's cell to the second site's cell and lastly the color of the bond.}
\label{tab:square_lattice}
\end{table}

\subsubsection{Square expansion}

\begin{figure}[h]
    \centering
    \includegraphics[width=0.7\linewidth]{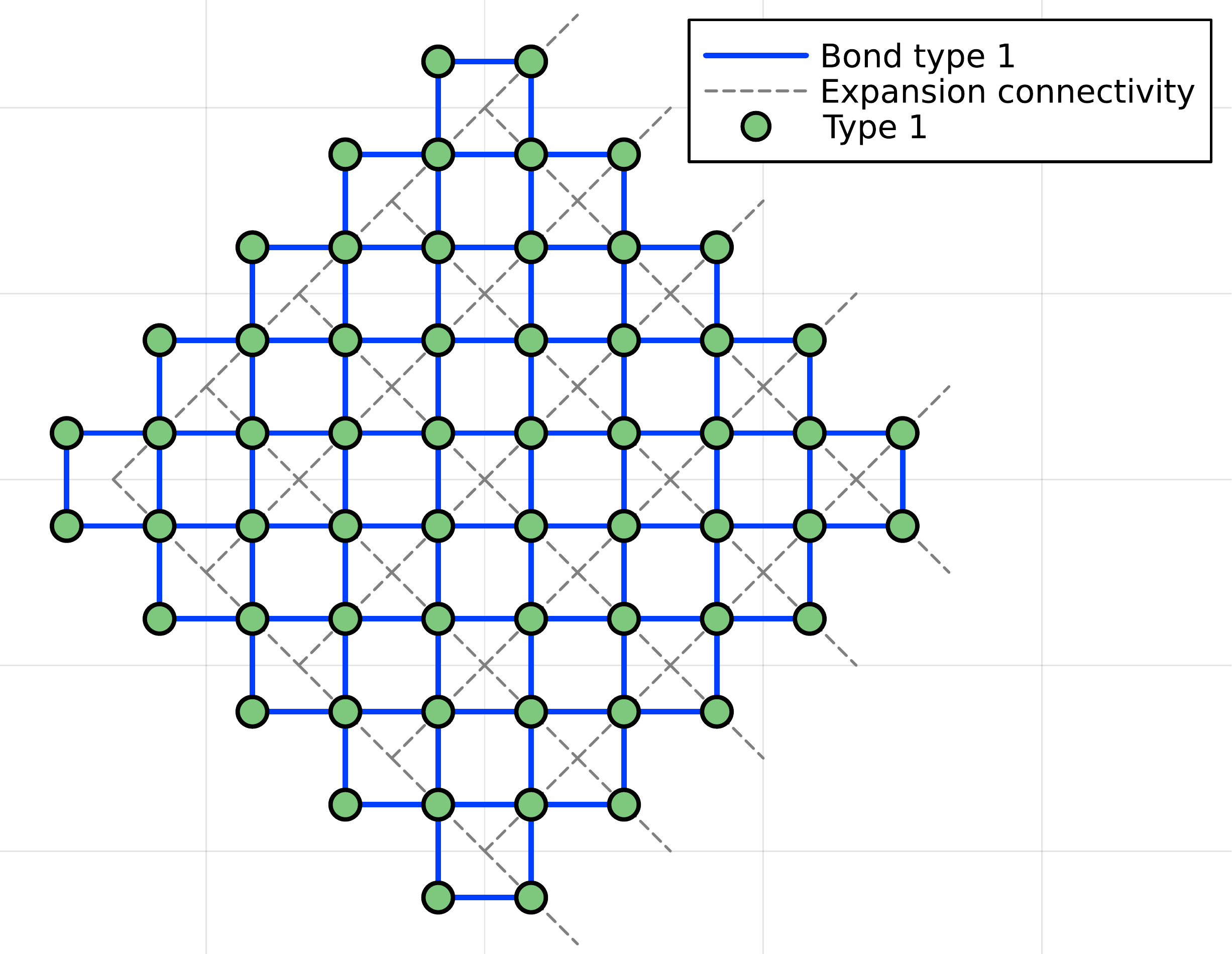}
    \caption{Visualization of the $2\times 2$ square expansion on the square lattice. The dashed gray lines indicate the underlying square lattice connecting the centers of the expansion units.}
    \label{fig:square}
\end{figure}

Table~\ref{tab:square_cluster} presents the input to \lincege for the square expansion on the square lattice and Fig.~\ref{fig:square} demonstrates the expansion units and the lattice used for the generation of clusters.

\begin{table}[h]
\centering
\begin{tabular}{@{}lll@{}}
\toprule
\textbf{Parameter} & \textbf{Symbol} & \textbf{Value} \\
\midrule
Primitive Vectors 
  & $p_1$ & $[1,\,0]$ \\
  & $p_2$ & $[0,\,1]$ \\
\addlinespace
Basis Vectors
  & $B$ & [[$b_1$, $b_2$, $b_3$, $b_4$]]\\
\addlinespace
  & $b_1$ & $[-\frac{1}{2}, -\frac{1}{2}]$ \\
  & $b_2$ & $[-\frac{1}{2}, \frac{1}{2}]$ \\
  & $b_3$ & $[\frac{1}{2}, -\frac{1}{2}]$ \\
  & $b_4$ & $[\frac{1}{2}, \frac{1}{2}]$ \\
\addlinespace
Basis Types 
  & $f$ & $[1,1,1,1]$ \\
\addlinespace
Lattice Bonds
  & $L_1$ & $[[1,1], [1,2], [0,0], 1]$\\
  & $L_2$ & $[[1,1], [1,3], [0,0], 1]$\\
  & $L_3$ & $[[1,2], [1,4], [0,0], 1]$\\
  & $L_4$ & $[[1,3], [1,4], [0,0], 1]$\\
\addlinespace
Expansion Bonds
  & $E_1$ & $[1,\,1,\,[1,0],\,1]$ \\
  & $E_2$ & $[1,\,1,\,[0,1],\,1]$ \\
\bottomrule
\end{tabular}
\caption{The input for the $2\times 2$ cluster expansion on the square lattice with uniform bond type. Since in a cluster expansion, one can have more than one expansion unit (e.g., up and down triangles on the Kagome lattice), basis vectors are organized as a list of lists ($B$). Each list in $B$ represents a different expansion unit. Here, the only expansion unit is a $2\times 2$ cluster with four bonds $b_1\dots b_4$. Lattice bonds and expansion bonds are also now separate inputs and represented differently to reflect the possibility of having more than one expansion unit. For example, the first element in $L_1$ ($[1,1]$) means the first site in the first expansion unit, identified by $b_1=B[1,1]$. The second element ($[1,2]$) is the second site across $L_1$, which is identified by $b_2=B[1,2]$. The third element in $L_1$ contains the integer coefficients of the translation vector required to get from the first site's cell to the second site's cell ($[0,0]$ in this case, which means $[0,0]\cdot [p_1,p_2]$), and the last element indicates the color/type of the bond. The expansion bonds are the bonds connecting expansion units. The first two integers are the expansion unit indices, both $1$ here since we have only one expansion unit. The third element contain the integer coefficients of the translation vector connecting the two expansion unit types, and the last integer again indicates that the bond type is $1$.}
\label{tab:square_cluster}
\end{table}

\subsection{Kagome Lattice}

\begin{figure}[h]
    \includegraphics[width=\linewidth]{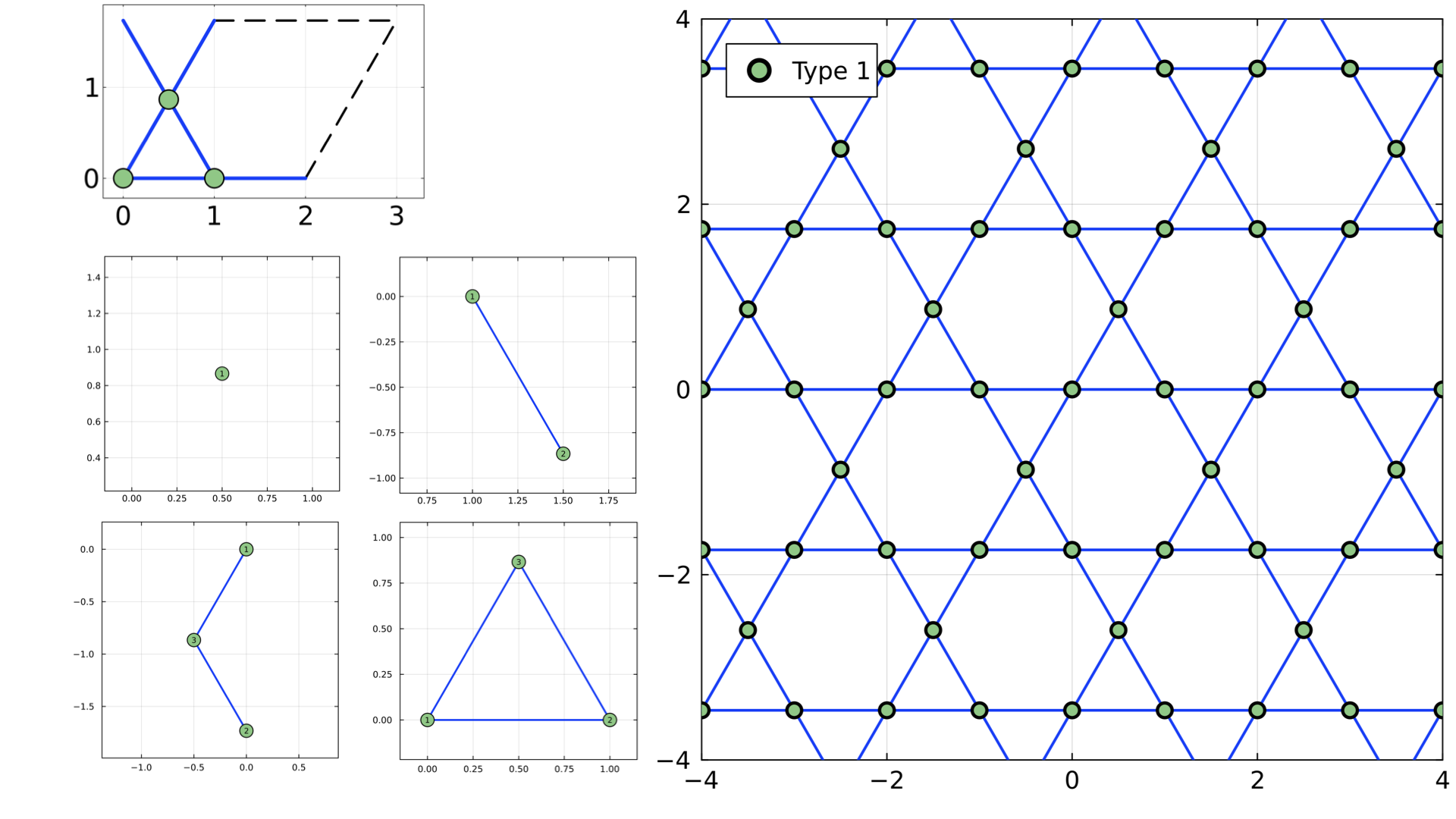}
    \caption{Kagome lattice unit cell, lattice plot and all topologically unique subclusters up to 3 sites. 
    }
    \label{fig:kagome}
\end{figure}

Table~\ref{tab:kagome} and Fig.~\ref{fig:kagome} present the input to \lincege for the Kagome lattice with a single site and bond type and the visualization of the unit cell and clusters in the series with up to three sites.

\begin{table}[h]
\centering
\begin{tabular}{@{}lll@{}}
\toprule
\textbf{Parameter} & \textbf{Symbol} & \textbf{Value} \\
\midrule
Primitive Vectors
  & $p_1$ & $[2,\,0]$ \\
  & $p_2$ & $[1,\,\sqrt{3}]$ \\
\addlinespace
Basis Vectors
  & $b_1$ & $[0,\,0]$ \\
  & $b_2$ & $[1,\,0]$ \\
  & $b_3$ & $[\frac{1}{2},\,\frac{\sqrt{3}}{2}]$ \\
\addlinespace
Basis Types
  & $f$ & $[1,\,1,\,1]$ \\
\addlinespace
Lattice/Expansion Bonds
  & $L_1$ & $[1,\,2,\,[0,0],\,1]$  \\
  & $L_2$ & $[2,\,3,\,[0,0],\,1]$ \\
  & $L_3$ & $[3,\,1,\,[0,0],\,1]$ \\
  & $L_4$ & $[1,\,2,\,[-1,0],\,1]$ \\
  & $L_5$ & $[1,\,3,\,[0,-1],\,1]$  \\
  & $L_6$ & $[2,\,3,\,[1,-1],\,1]$ \\
\bottomrule
\end{tabular}
\caption{The input for the site expansion on the Kagome lattice with a three-atom basis and a uniform bond type.}
\label{tab:kagome}
\end{table}

\subsection{Shastry-Sutherland Lattice}

\begin{figure}[h]
    \includegraphics[width=\linewidth]{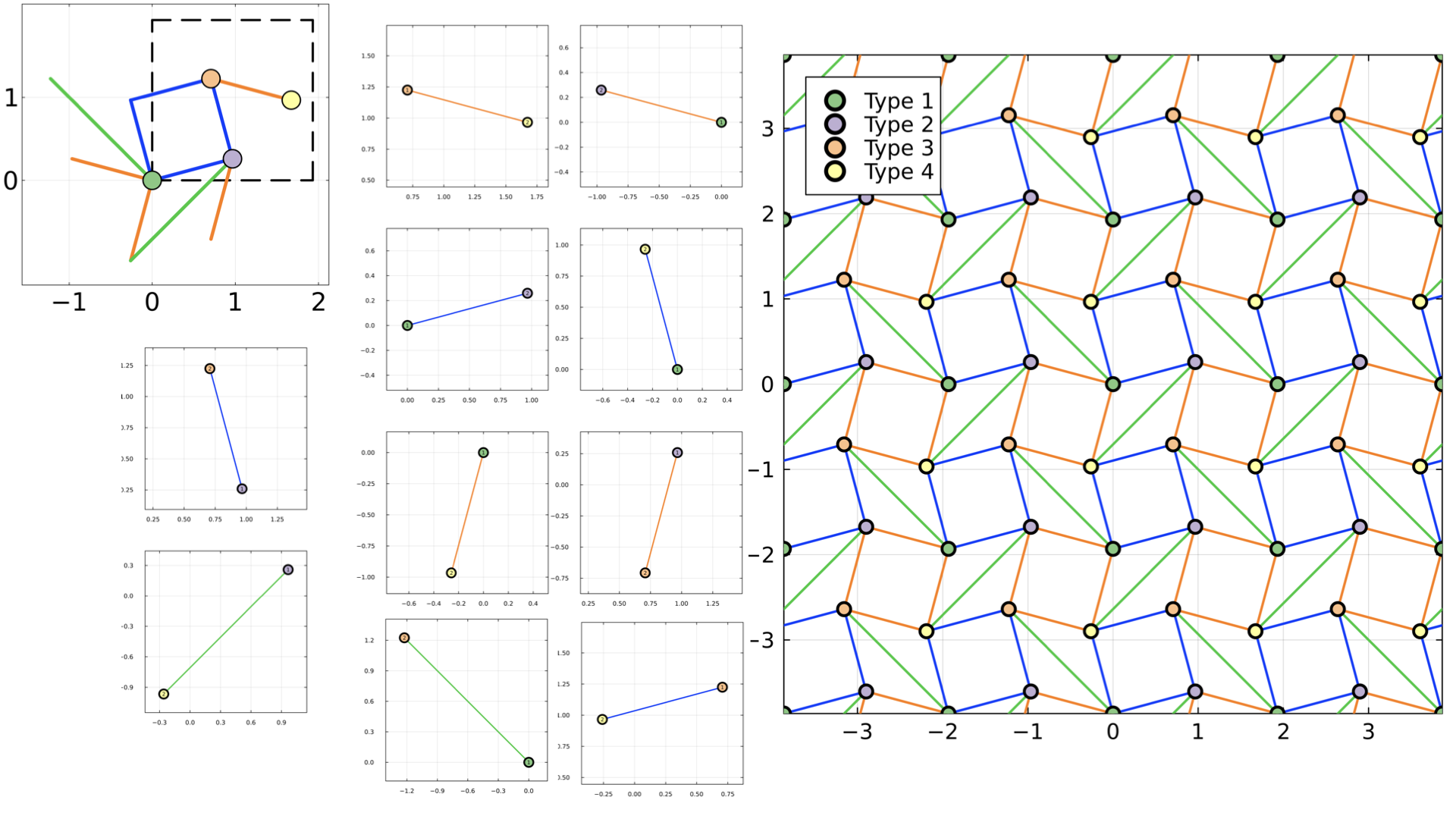}
    \caption{Shastry Sutherland lattice unit cell, lattice plot and all two site topologically unique subclusters. 
    }
    \label{fig:shastry_sutherland}
\end{figure}

Table~\ref{tab:shastry_sutherland} and Fig.~\ref{fig:shastry_sutherland} present the input to \lincege for an isotropic Shastry-Sutherland lattice with four different types of sites and four different types of bonds, along with the visualization of the unit cell and clusters with up to two sites (dimers with different combinations of site and bond types).

\begin{table}[h]
\centering
\begin{tabular}{@{}llll@{}}
\toprule
\textbf{Parameter} & \textbf{Symbol} & \textbf{Value}\\
\midrule
Primitive Vectors
  & $p_1$ & $\bigl[\frac{\sqrt{3}+1}{\sqrt{2}},\,0\bigr]$ & \\
  & $p_2$ & $\bigl[0,\,\frac{\sqrt{3}+1}{\sqrt{2}}\bigr]$ & \\
\addlinespace
Basis Vectors
  & $b_1$ & $[0,\,0]$                                                                              \\
  & $b_2$ & $\bigl[\frac{\sqrt{3}+1}{2\sqrt{2}},\,\frac{\sqrt{3}-1}{2\sqrt{2}}\bigr]$             \\
  & $b_3$ & $\bigl[\frac{\sqrt{2}}{2},\,\frac{\sqrt{6}}{2}\bigr]$                                 \\
  & $b_4$ & $\bigl[\frac{\sqrt{3}+3}{2\sqrt{2}},\,\frac{\sqrt{3}+1}{2\sqrt{2}}\bigr]$             \\
\addlinespace
Basis Types
  & $f$ & $[1,\,2,\,3,\,4]$ & \\
\addlinespace
Lattice/Expansion Bonds
  & $L_1$    & $[1,\,2,\,[0,0],\,1]$    \\
  & $L_2$    & $[2,\,3,\,[0,0],\,1]$    \\
  & $L_3$    & $[1,\,4,\,[-1,0],\,1]$  \\
  & $L_4$    & $[3,\,4,\,[-1,0],\,1]$   \\

  & $L_5$    & $[3,\,4,\,[0,0],\,2]$    \\
  & $L_6$    & $[2,\,3,\,[0,-1],\,2]$   \\
  & $L_7$    & $[1,\,2,\,[-1,0],\,2]$   \\
  & $L_8$    & $[1,\,4,\,[-1,1],\,2]$   \\

  & $L_9$    & $[2,\,4,\,[-1,-1],\,3]$  \\
  & $L_{10}$ & $[1,\,3,\,[-1,0],\,3]$   \\
\bottomrule
\end{tabular}
\caption{The input for the site expansion on the Shastry-Sutherland lattice with a multi-type four-site basis and three distinct bond types.}
\label{tab:shastry_sutherland}
\end{table}

\clearpage
\section{Generated Tables for various lattice geometries}
\label{appendix:counts}
\begin{table}[!h]
 \centering
\begin{tabular}{rrrrrr}
  \hline
  Order & No. of isomorphic clusters & No. of symmetric clusters & $\sum L(c)$ & $\sum S(c)$ & $\sum F(c)$ \\
  \hline
  1 & 1 & 1 & 1.0 & 0 & 1.0 \\
  2 & 1 & 1 & 2.0 & 2 & -2.0 \\
  3 & 1 & 2 & 6.0 & 5 & 0.0 \\
  4 & 3 & 5 & 19.0 & 31 & 1.0 \\
  5 & 4 & 12 & 63.0 & 69 & 0.0 \\
  6 & 10 & 35 & 216.0 & 289 & 0.0 \\
  7 & 19 & 108 & 760.0 & 848 & 0.0 \\
  8 & 51 & 369 & 2725.0 & 3577 & 1.0 \\
  9 & 112 & 1285 & 9910.0 & 11849 & -1.0 \\
  10 & 300 & 4655 & 36446.0 & 48071 & 2.0 \\
  11 & 746 & 17073 & 135268.0 & 178158 & -4.0 \\
  12 & 2042 & 63600 & 505861.0 & 730398 & 11.0 \\
  \hline
\end{tabular}
\caption{Automatically generated table of the cluster counts on an infinite square lattice using the site expansion. The first column is the order, which is equivalent to the number of sites. The second and third columns are the number of isomorphic and symmetric clusters, respectively. The final three columns are more complex cluster counts. $\sum L(c)$ is the sum of the lattice constants of all clusters in a given order. $\sum S(c)$ is the total number of subclusters of the isomorphically distinct clusters in that order. $\sum F(c)$ is the sum over the final coefficients after the recursive summation for each cluster up to that order is performend.}\label{tab:square_lattice_infinite}
\end{table}

\begin{table}[!h]
 \centering
\begin{tabular}{rrrrr}
  \hline
  Order & No. of isomorphic clusters & $\sum L(c)$ & $\sum S(c)$ & $\sum F(c)$ \\
  \hline
  1 & 1 & 1.0 & 0 & 1.0 \\
  2 & 1 & 3.0 & 2 & -3.0 \\
  3 & 2 & 11.0 & 11 & 2.0 \\
  4 & 4 & 44.0 & 43 & 0.0 \\
  5 & 8 & 186.0 & 154 & 0.0 \\
  6 & 22 & 814.0 & 706 & 1.0 \\
  7 & 54 & 3652.0 & 2793 & -1.0 \\
  8 & 156 & 16689.0 & 12731 & 3.0 \\
  9 & 457 & 77359.0 & 58339 & -4.0 \\
  10 & 1424 & 362671.0 & 282101 & 9.0 \\
  11 & 4505 & 1.71603e6 & 1377550 & -15.0 \\
  12 & 14791 & 8.18221e6 & 6971412 & 30.0 \\
  \hline
\end{tabular}
\caption{Same as Tab.~\ref{tab:square_lattice_infinite} but for the infinite triangular lattice.}
\end{table}

\begin{table}[!h]
 \centering
\begin{tabular}{rrrrr}
  \hline
  Order & No. of isomorphic clusters & $\sum L(c)$ & $\sum S(c)$ & $\sum F(c)$ \\
  \hline
  1 & 1 & 1.0 & 0 & 1.0 \\
  2 & 1 & 2.0 & 2 & -2.0 \\
  3 & 2 & 4.66667 & 11 & 0.666667 \\
  4 & 2 & 12.0 & 20 & 0.0 \\
  5 & 4 & 33.0 & 70 & 0.0 \\
  6 & 7 & 93.6667 & 189 & 0.333333 \\
  7 & 12 & 272.0 & 494 & 0.0 \\
  8 & 22 & 805.0 & 1369 & 0.0 \\
  9 & 45 & 2420.0 & 4117 & 0.0 \\
  10 & 88 & 7358.0 & 11759 & 1.0 \\
  11 & 183 & 22581.0 & 35210 & -1.0 \\
  12 & 389 & 69850.7 & 108454 & 0.666667 \\
  \hline
\end{tabular}
\caption{Same as Tab.~\ref{tab:square_lattice_infinite} but for the infinite kagome lattice.}
\end{table}

\begin{table}[!h]
 \centering
\begin{tabular}{rrrrr}
  \hline
  Order & No. of isomorphic clusters & $\sum L(c)$ & $\sum S(c)$ & $\sum F(c)$ \\
  \hline
  0 & 1 & 1.0 & 0 & 1.0 \\
  1 & 1 & 0.5 & 4 & -1.5 \\
  2 & 1 & 1.0 & 9 & 0.0 \\
  3 & 2 & 3.0 & 30 & 0.0 \\
  4 & 5 & 9.5 & 113 & 0.5 \\
  5 & 11 & 31.5 & 346 & 0.0 \\
  6 & 31 & 108.0 & 1352 & 0.0 \\
  7 & 86 & 380.0 & 4979 & 0.0 \\
  8 & 272 & 1362.5 & 21217 & 0.5 \\
  9 & 850 & 4955.0 & 87278 & -0.5 \\
  10 & 2814 & 18223.0 & 384569 & 1.0 \\
  11 & 9364 & 67634.0 & 1691315 & -2.0 \\
  12 & 31913 & 2.5293e5 & 7673523 & 5.5 \\
  \hline
\end{tabular}
\caption{Same as Tab.~\ref{tab:square_lattice_infinite} but for the infinite square lattice corner sharing square expansion.}
\end{table}

\begin{table}[!h]
\centering
\begin{tabular}{rrrrr}
  \hline
  Order & No. of isomorphic clusters & $\sum L(c)$ & $\sum S(c)$ & $\sum F(c)$ \\
  \hline
  0 & 1 & 1.0 & 0 & 1.0 \\
  1 & 1 & 0.25 & 4 & -0.75 \\
  2 & 1 & 1.5 & 10 & -1.5 \\
  3 & 3 & 12.5 & 53 & 2.0 \\
  4 & 8 & 118.75 & 222 & 0.25 \\
  5 & 25 & 1220.25 & 1030 & -1.5 \\
  6 & 100 & 13208.8 & 6145 & 7.5 \\
  7 & 466 & 1.48346e5 & 43008 & -1.0 \\
  \hline
\end{tabular}
\caption{Same as Tab.~\ref{tab:square_lattice_infinite} but for the infinite pyrochlore lattice unit cell expansion.}
\end{table}

\end{appendices}

\end{document}